# Scheduling for Flexible Manufacturing System with Objective Function to be Minimization of Total Processing Time and Unbalance of Machine Load

U Yong Nam[1], Ri Taehyong[1,2*]

[1]Institute of Robot Engineering, Kim Chaek University of Technology

[2]College of Information and Control Engineering, China University of Petroleum (East China), Qingdao, 266580, China

*corresponding author: lb1405001@s.upc.edu.cn

**Abstract**: For scheduling in flexible manufacturing system (FMS), many factors should be considered, it is difficult to solve the scheduling problem by satisfying different criteria (production cost, utilization of system, number of movements of part, make-span, and tardiness in due date and so on) and constrains. The paper proposes mathematical model of a job shop scheduling problem (JSSP) to balance the load of all machines and utilize effectively all machines in FMS. This paper defines the evaluation function of the unbalance of the machine load and formulates the optimization problem with two objectives minimizing unbalance of the machine load and the total processing time, scheduling problem having been solved by integer linear programming, thus scheduling problem having been solved. The results of calculation show that the total processing time on all machines is reduced and machine loading is balanced better than previous works, and job shop scheduling also could be scheduled more easily in FMS.

## 1. Introduction

FMS, which is designed by combining the high efficiency of mass production with the flexibility of job shop in manufacturing system, is used for the low volume of multi-variety products and allow processing a variety of products at the same time. An FMS consists of many resources such as CNC machine tools, automated material handling system, robots, and in-process storage facilities. The most important thing in a job shop scheduling problem (JSSP) of FMS is to find a way to assign the shared resources and to find sequence of jobs so that the evaluation criteria, such as production cost, utilization of system, number of movements of parts, make-span, balancing in machine load, and tardiness in due date, can be minimized (maximized) and production constraints can be satisfied [1]. For studying JSSP, therefore, all of constraints are not satisfied, but could select one or a few evaluation criterions under a certain production environment. JSSP, having to solve in the efficient running of FMS, is divided into five sub-problems by Stecke (1985), namely part type selection, machine loading, machine grouping, production ratio, and resource allocation problems. However, the unbalanced reason in part selection and machine loading problems may be the fact that the values of machining parameters are predetermined and but not optimized in FMS planning process. Although the installation of FMS requires a greater capital investment, their expected benefits, which include increased machine utilization, less machines, reduced floor space, greater responsiveness to changes, reduced inventories, lower manufacturing lead times, and higher labour productivity, are substantial. Mussa I. Mgwatu [3] made an attempt to integrate the decisions of part selection, machine loading and machining optimization problems for a more balanced workload and effective FMS. Ulrich A.W. *et al.* [4] proposed several nonlinear optimization models in order to optimize the allocation of workloads between a job shop and an FMS and illustrated that the models allow optimizing performance parameters like throughput, work-in-process inventory, utilization, and production lead time. Most of the previous research, used the objectives of minimize processing cost, total processing time, frequency of tool movement, or inventory cost for loading and scheduling models and some researchers have used the minimization of the make-span as their objective. Mansour Abou Gamila *et al*. [5] formulated the scheduling problem as a zero-one integer programming problem to load and route the operations and tools between machines and based on the results developed heuristics to solve the operations scheduling. Mohit Goswami *et al*. [6] solved the sub-problems like tool-part grouping, job allocation on machines and minimization of make-span in a FMS. They also performed Tool and part grouping using "principal component analysis", and carried out tool allocation using priority-based approach by developing a potency index. And then, the concept has evolved based on threshold machining time, minimum operation time, and maximum possible operation time to minimize the make-span of the system. Xiao ming ZHANG *et al*. [7] proposed a part-centric tool allocation strategy and a tool-centric tool allocation strategy and then evaluated their performances using simulations to compare their characteristics under different production conditions. S. Rahimifard *et al*. [8] provided a novel machine loading policy in flexible machining cell which meet the delivery dates of production orders and reduce the manufacturing cost at the same time. Chinyao Low *et al*.[9] have used for a multi-objective model with consideration of minimum mean job flow time, mean job tardiness, and minimum mean machine idle time, simultaneously to solve the scheduling problems in FMS. L.J. Zeballos *et al*. [10] proposed constraint programming model for the scheduling of FMSs with machine and tool limitations. F. Jolai *et al*.[11] considered a bi-objective no-wait two-stage flexible flow shop with the objectives of minimizing make-span and maximum tardiness of jobs.

Though many objective functions have been discussed in literary review, those criteria used in their research could not avoid problems induced by unbalance of machine load and ineffective machine utilization. In addition, there are some different opinions about determining the machine loads quantitatively.

The aim of the present work is to construct mathematical model of a job shop scheduling problem (JSSP) to balance the load of all machines and utilize effectively all machines in FMS.

The rest of the paper is organized as follows. Section 2 describes the rules of a job shop production and mathematical models for evaluating the unbalance of the machine loads in FMS and formulates an optimal problem with objective functions minimizing the total processing time and unbalance of the machine load for scheduling in FMS. Section 3, in order to solve optimal problem by integer linear programming, proposes the treatment method of absolute value and applies the proposed model to a given example. Section 4 presents the calculated results of application and discussion and section 5 describes the conclusion of this paper.

**2. Problem statement**

2.1. Assumptions and notations

2.1.1. Assumptions

i. The individual tool change time is not considered and the storage capacity of tool in every machine is enough.
ii. We consider an FMS to consist of functionally dissimilar CNC machines.
iii. A part can be processed on more than one machine and it is not restricted to only one machine.
iv. Each machine can process only one part at a time and each part can be processed on only one machine at a time, and each operation cannot be interrupted.
v. The movement time of part is neglected.
vi. Running an FMS, all machines operate normally without troubles.

2.1.2. Notations

The scheduling problem of FMS proposed in this paper is described as follows: when given a set of machines $K = \{1, 2, \cdots, m\}$ and a set of parts $P = \{1, 2, \cdots, N\}$ in an FMS system, each part $i$ ($i$=1, 2, ..., $N$,) consists of a sequence of $p(i)$ operations and has its corresponding due date $D_i$. Each operation $J(i)$ of part $i$ ($J(i)$=1, 2, ..., $p(i)$) can be processed on an arbitrary machine $m$ without interrupt. The setup time is assumed to be separate from its corresponding processing time. Thus, the setup task can be started in advance when the particular machine is free and consequently, the flow time of the parts can be shortened. Each machine $m$ ($m$=1, 2, ..., $M$,) can process at most one operation at a time, and at most one operation of each part $P_i$ can be processed at a time.

The present paper aims to determine the operation scheduling in FMS that minimize the processing time on every machine, while balancing of all machine load.

Two main objectives are:

i. Minimization of the total processing time.

ii. Minimization of the evaluating function for the unbalance of machine load.

The notations used for this paper are as follows:

| | | |
|---|---|---|
| $i$ | Part | $i$=1, 2,..., $N$ |
| $J(i)$ | Operation of a part $i$ | $J(i)$=1, 2,..., $p(i)$ |
| $m, n$ | Machine | $m, n$=1, 2, ..., $M$ |
| $l$ | Tool | $l$=1, 2,..., $T$ |
| $A$ | Machining stage | $A$=1, 2, ..., $z$ |
| $TL_l$ | Life of tool $l$ | |
| $MT_{i, J(i), m, l}$ | Processing time for operation $J(i)$ of part $i$ using tool $l$ on machine $m$ | |
| $S$ | The maximum completion time | |
| $OC_{i, J(i), m, l}$ | Cost of operation $J(i)$ of part $i$ using tool $l$ on machine $m$ | |
| $C$ | Total target cost for the processing | |
| $N_i$ | Number of movements of part $i$ between machines | |

$SC_i$     Setup cost for part $i$

$D_i$     Due date of part $i$

$SC$     Limit on setup cost

$S_m$     Tool magazine capacity of machine $m$

$WL_m, WL_n$     Machine load of machine $m, n$

$L$     Evaluation function of machine load

$F_1$     Objective for total processing time

$F_2$     Objective for machine load ($F_2 = L$)

$z_i$,     variables of unconditional optimal problem

$\Delta_{ij}$     Deviation of absolute objective

$\delta_{ij}^+$     Positive between unconditional optimal variables

$\delta_{ij}^-$     Negative between unconditional optimal variables

$y_i$     Order number to expression $|WL_m - WL_n|$

$\bar{x}, \bar{\delta}^+, \bar{\delta}^-$     Optimal solution of the unconditional optimal problem

$W_1, W_2$     Weight factors of objectives

*Decision variables*

$X_{m, l}$     Zero-one variable, equal 1 if tool $l$ is assigned to machine $m$; equal to 0 otherwise

$x_{i, J(i), m, l, A}$     Zero-one variable, equal 1 if operation $J(i)$ of part $i$ is assigned to machine $m$ with tool $l$; equal 0 otherwise

2.2. The evaluation function for unbalance of machine load in system

Considering the machine load for scheduling of FMS aims to balance the utilization of every machine when accomplish one batch of parts on every machine in system and to reduce mean stream time for a batch.

When all parts will be accomplished in FMS consisting of *m* machines, the loading of machine *m*, as summed up the processing time of all operations on machine *m*, is calculated as follows:

$$WL_m = \sum_{A=1}^{z} \sum_{J(i)=1}^{p(i)} \sum_{l=1}^{T} MT_{i,J(i),m,l} x_{i,J(i),m,l,A} . \qquad (1)$$

In this paper the evaluation function for the unbalance of machine load in system is defined as follows:

$$L = \sum_{m=1}^{M} \sum_{n=m+1}^{M} |WL_m - WL_n| . \qquad (2)$$

The smaller the value of Eq. (2), the better the loading between machines is balanced.

2.3. Optimization

The main purpose of optimization in this paper is to minimize the evaluation function for unbalance of the machine load and the total processing time for all parts in system.

2.3.1. Objective functions

The total processing time, required to complete all parts, is calculated as follows:

$$F_1 = \sum_{A=1}^{z} \sum_{m=1}^{M} \sum_{l=1}^{T} \sum_{i=1}^{N} \sum_{J(i)=1}^{p(i)} MT_{i,J(i),m,l} \times x_{i,J(i),m,l,A} . \qquad (3)$$

From Eq. (2), the second objective is calculated as follows:

$$F_2 = \sum_{m=1}^{M} \sum_{n=m+1}^{M} |WL_m - WL_n| . \qquad (4)$$

### 2.3.2. Constraints

Every operation is processed using a suitable tool and available machine.

$$\sum_{m=1}^{M}\sum_{l=1}^{T} x_{i,J(i),m,l,A} = 1, \quad \forall J(i), i, A \tag{5}$$

The completion time for every machine cannot exceed the maximum completion time.

$$\sum_{A=1}^{z}\sum_{l=1}^{T}\sum_{i=1}^{N}\sum_{J(i)=1}^{p(i)} MT_{i,J(i),m,l} \times x_{i,J(i),m,l,A} \leq S, \quad \forall m \tag{6}$$

The completion time for every part cannot exceed its due date.

$$\sum_{A=1}^{z}\sum_{m=1}^{M}\sum_{L=1}^{T}\sum_{J(i)=1}^{p(i)} MT_{i,J(i),m,l} \times x_{i,J(i),m,l,A} \leq D_i, \quad \forall i \tag{7}$$

Each tool is assigned to only one machine.

$$\sum_{m=1}^{M} x_{m,l} \leq 1, \quad \forall l \tag{8}$$

All operations cannot exceed the target cost.

$$\sum_{A=1}^{z}\sum_{m=1}^{M}\sum_{l=1}^{T}\sum_{i=1}^{N}\sum_{J(i)=1}^{p(i)} OC_{i,J(i),m,l} \times x_{i,J(i),m,l,A} \leq C \tag{9}$$

The setup cost of parts on machines cannot exceed allowed setup cost.

$$\sum_{i=1}^{N} N_i \times C_i \leq SC \tag{10}$$

Where $N_i$ is calculated as follows:

$$N_i = \sum_{m=1}^{M} \left| x_{i,J(i),m,l,A} - x_{i,J(i)+1,m,l,A} \right| / 2, \quad \forall J(i), i, l, A. \tag{11}$$

## 3. Solving by integer linear programming

### 3.1. Treatment of objective function

The evaluation function for unbalance of machine load of two objective functions cannot be solved by linear programming. Therefore, the absolute value of function should be changed so that it can be solved by linear programming.

Consider the following unconditional optimal problem.

$$\min \sum_{\substack{i,j=1 \\ i \neq j}}^{l} |z_i - z_j| \tag{12}$$

Then, the concept of deviation can be defined as follows:

$$\Delta_{ij} = |z_i - z_j|, \ i, j = 1, 2, l \tag{13}$$

Where

$$\text{Positive deviation-} \delta_{ij}^+ = \begin{cases} z_i - z_j & , z_i \geq z_j \\ 0 & , z_i < z_j \end{cases}, \ i, j = 1, 2, l$$

$$\text{Negative deviation-} \delta_{ij}^- = \begin{cases} 0 & , z_i \geq z_j \\ z_i - z_j & , z_i < z_j \end{cases}, \ i, j = 1, 2, l$$

And then the relation between positive deviation ($\delta_{ij}^+$) and negative deviation ($\delta_{ij}^-$) is as follows:

$$\begin{cases} \delta_{ij}^+ + \delta_{ij}^- = \Delta_{ij} \\ \delta_{ij}^+ - \delta_{ij}^- = z_j - z_j \\ \delta_{ij}^+ \cdot \delta_{ij}^- = 0, \ \delta_{ij}^+, \delta_{ij}^- \geq 0, \ i, j = 1, 2, \cdots, l \end{cases} \tag{14}$$

On the basis of above consideration Eq. (13) can be replaced as follows:

$$\begin{cases} \min \sum_{\substack{i,j=1 \\ i \neq j}}^{l} (\delta_{ij}^+ + \delta_{ij}^-), \\ z_i - \delta_{ij}^+ + \delta_{ij}^- = z_j, \\ \delta_{ij}^+ \cdot \delta_{ij}^- = 0, \\ \delta_{ij}^+, \delta_{ij}^- \geq 0, \quad i,j = 1,2,\cdots,l \end{cases} \quad (15)$$

Removing constrain $\delta_{ij}^+ \cdot \delta_{ij}^- = 0$ from the above expression, Eq. (16) is as follows:

$$\begin{cases} \min \sum_{\substack{i,j=1 \\ i \neq j}}^{l} (\delta_{ij}^+ + \delta_{ij}^-), \\ z_i - \delta_{ij}^+ + \delta_{ij}^- = z_j \\ \delta_{ij}^+, \delta_{ij}^- \geq 0, \quad i,j = 1,2,\cdots,l \end{cases} \quad (16)$$

And then, there is the following theorem that shows the relation between Eq. (12) and (16).

[**Theorem**] Suppose that $(\bar{x}, \bar{\delta}^+, \bar{\delta}^-)$ are the optimal solution of Eq. (16), then $\bar{x}$ is always the optimal solution of Eq. (12), where $\bar{\delta}^+ = (\bar{\delta}_1^+, \bar{\delta}_2^+, \cdots, \bar{\delta}_l^+)$, $\bar{\delta}^- = (\bar{\delta}_1^-, \bar{\delta}_2^-, \cdots, \bar{\delta}_l^-)$ [12].

(Proof cut out.)

3.2. Objective function for optimization

In this paper, two objective functions, including minimizing of the total processing time and minimizing of the evaluation function for unbalance of machine load, are considered. Therefore a weight between 0 and 1 is allocated to both objective functions, which the sum of weights is 1. Then, two objectives are summed together and one objective is made of:

$$\min Z = W_1 \times F_1 + W_2 \times F_2 \quad (17)$$

$$W_1 + W_2 = 1, W_1, W_2 \geq 0$$

Where $F_1$ is total processing time, $F_2 = L = \sum_{m=1}^{M} \sum_{n=m+1}^{M} |WL_m - WL_n| = \sum_{i=1}^{M \cdot (M-1)} y_i$ is the evaluation function for unbalance of machine load, $y_i$ are variable that give a sequence number to expression $|WL_m - WL_n|$ in order and $W_1$, $W_2$ are 0.5 respectively.

## 4. Results and discussion

In order to examine the efficiency of the proposed model, we applied the model to the first example, for which Mansour Abou Gamila *et al*. and Saeid Motavalli [5] used. The example consists of four parts, each with four operations. There are 4 machines in the FMS and each machine has a tool magazine with a capacity of 40 tool slots. There are 20 types of tools available for processing all parts. Table 1 shows the set up costs and the due dates for each part. Table 2 shows the operation times and the associated machining costs for performing an operation on a part on each machine using a particular tool. The tool life for each tool is 150 min. The total target cost is $4,600. The limit on set up cost is $700.

Table 1 . Due date and set up costs for parts

|  | Part1 | Part2 | Part3 | Part4 |
|---|---|---|---|---|
| Due date | 380 | 420 | 350 | 400 |
| Set up cost | 90 | 70 | 140 | 110 |

Table 2 . Machining times and costs

| Part type | 1 | | | | 2 | | | |
|---|---|---|---|---|---|---|---|---|
| Operation sequence | 1 | 2 | 3 | 4 | 1 | 2 | 3 | 4 |
| Compatible tool No. | 1  2 | 4  7 | 6 | 10  13 | 1  3 | 8  6 | 10  17 | 4  12 |
| Processing each Machine,(min) M1 | 104 | 68 | 84 | 114  114 | | 25  106 | | 96 |

|  |  | 1 | 2 | 3 | 4 | 1 | 2 | 3 | 4 |
|---|---|---|---|---|---|---|---|---|---|
|  | M2 | 110 | 120 130 | 110 | 76 | 126 98 |  | 66 | 116 |
|  | M3 |  | 101 106 | 118 |  |  | 119 | 29 112 | 84 |
|  | M4 |  |  |  | 100 |  |  |  |  |
| Machining cost of each operation(*j*) of each of part(*i*) using tool (*l*) on machining | M1 | 24 |  | 14 |  | 21 | 21 28 | 54 27 | 17 |
|  | M2 | 23 |  | 35 29 | 40 | 35 | 35 21 | 19 | 14 |
|  | M3 |  | 14 30 |  | 40 |  | 33 | 42 25 | 34 |
|  | M4 |  |  |  | 60 |  |  |  |  |

| Part type | 3 | | | | 4 | | | |
|---|---|---|---|---|---|---|---|---|
| Operation sequence | 1 | 2 | 3 | 4 | 1 | 2 | 3 | 4 |
| Compatible tool No. | 12 15 | 9 18 | 11 19 | 3 14 | 2 4 | 5 20 | 13 14 | 5 8 |

| | | 1 | 2 | 3 | 4 | 1 | 2 | 3 | 4 |
|---|---|---|---|---|---|---|---|---|---|
| Processing each Machine, (min) | M1 | 67 |  | 82 |  | 137 |  | 68 |  |
|  | M2 |  | 117 47 | 85 110 |  | 114 | 38 | 115 53 |  |
|  | M3 | 102 | 90 |  |  | 49 | 140 |  | 87 |
|  | M4 |  | 134 120 | 40 | 132 |  | 118 | 120 |  |
| Machining cost of each operation(*j*) of each of part(*i*) using tool (*l*) on machining | M1 | 10 |  |  |  | 36 |  | 13 |  |
|  | M2 |  | 17 27 | 17 19 |  | 31 | 39 | 41 |  |
|  | M3 | 31 | 25 |  |  | 22 | 37 |  | 24 |
|  | M4 |  | 14 25 | 40 | 22 |  | 29 | 35 |  |

Fig.1-Fig.3, Table 3 and 4 show the results achieved by applying the proposed model. Fig.1 shows Gantt chart for scheduling of operations on the assigned machines, Fig.2 and Fig.3 show utilization and processing time of every machine. As shown in Fig. 2, 3 and table 3, 4, using the proposed model resulted in:

i. The total processing time was 1,193 min; it was reduced less than previous works and total cost was 4390$, it was not exceed total target cost (Table 3).

ii. The maximum completion time was 435 min; it was increased more than M. A. Gamila and S. Motavalli's 403min, but reduced less than Sarin and Chen's 558 min (Table 4).

iii. The mean processing time was 298.25 min; it was reduced less than previous works (Fig.4).

iv. The maximum deviation in machine load was reduced less than Sarin and Chen's: 9 versus 65(Table 4).

v. The utilizations of all the machines were in 0.68-0.7% over the maximum completion time (435min) and then it was not much different in all machines (Fig.2, 3 and table 4).

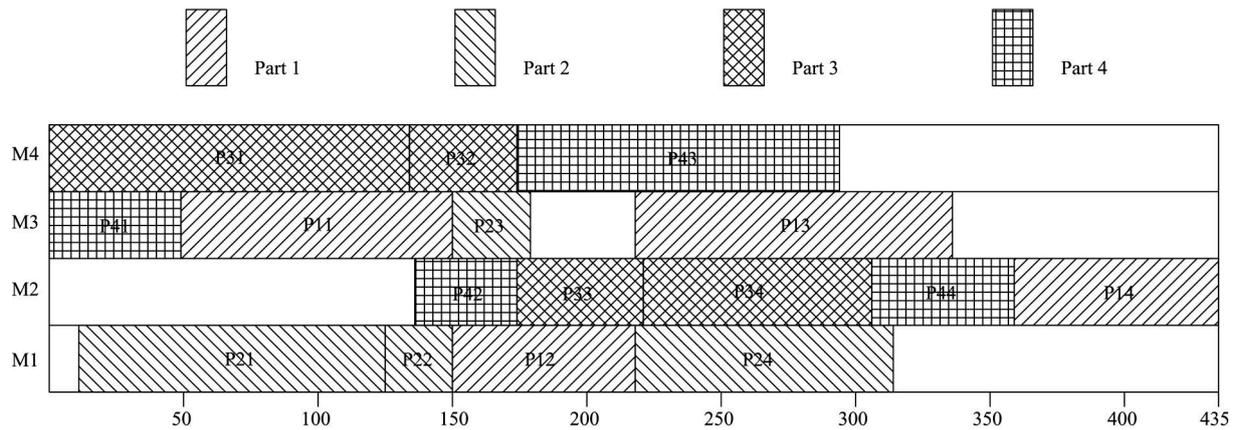

Fig.1. Gantt chart

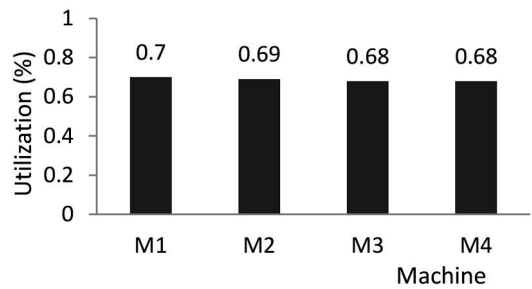

Fig.2. Utilization of machines

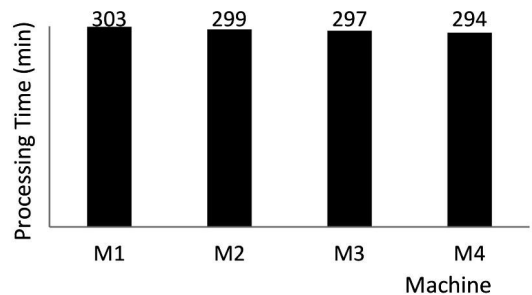

Fig.3. Processing time of machines
13

| Machines | Operation assigned | Tools assigned | Processing time | Completion time | Processing cost | Utilization |
| --- | --- | --- | --- | --- | --- | --- |

| | | | | | |
|---|---|---|---|---|---|
| M1 | 21,22,12, 24 | 1,7,12,16 | 303 | 314 | 1060 | 0.70 |
| M2 | 42,14,33,34 | 3,5,10,19,20 | 299 | 435 | 1260 | 0.69 |
| M3 | 11,41,23,13,44 | 2,6,17 | 297 | 336 | 1180 | 0.68 |
| M4 | 31,32,43 | 13,15,18 | 294 | 294 | 890 | 0.68 |
| Sum | | | 1193 | - | 4390 | 0.688 |

Table 3. Assignment of tools and operations to machines and the utilization of machines

Table 4. Assignment of tools and operations to machines and the utilization of machines

| Model | Processing cost | Total time | Max. completion time | Mean operating time of machine | Max. deviation of machine load |
|---|---|---|---|---|---|
| Sarin and Chen [5] | 3590 | 1369 | 558 | 394 | - |
| M.A.Gamila and S.Motavalli[5] | 4540 | 1201 | 403 | 300.25 | 65 |
| Our Model | 4390 | 1193 | 435 | 298.25 | 9 |

As seen in the result of calculation, total processing time was reduced to 1193 which is less than previous ones and the maximum deviation was reduced to one-seventh of Sarin and Chen. And difference in utilization of machines is 0.02, which is very little. This is because the minimization of the total processing time and evaluation function for unbalance of machine load were used as an objective function. The proposed model, therefore, can be used to balance machine load and improve utilization of machine for a JSSP in an FMS consisting of the functionally similar CNC machines. However, when FMS includes a specific machine which is specially designed for a high productivity or specialty, it might be considered again.

5. Conclusion

This paper proposed the evaluation function for unbalance of machine load in a system and described the method for scheduling of FMS that satisfies two objectives minimizing the total processing time of all products and unbalance of machine load. Using the proposed model could balance the load of all machines in system, could utilize effectively all machines, and also could simplify the scheduling for FMS. If the proposed model is combined with the artificial intelligent approaches, it could obtain better results.

## References

1. B.B. Choudhury, B.B. Biswal, D. Mishra, R. N, Mahapatra. Appropriate Evolutionary Algorithm for Scheduling in FMS, 2009 World Congress on Nature & Biologically Inspired Computing (NaBIC 2009) 1139-1144

2. Zubair M. Mohamed, Ashok Kumar, Jaideep Motwani, An improved part grouping model for minimizing makespan in FMS, European Journal of Operational Research 116 (1999) 171-182

3. Mussa I. Mgwatu, Integration of part selection, machine loading and machining optimisation decisions for balanced workload in flexible manufacturing system, International Journal of Industrial Engineering Computations 2 (2011) 913–930

4. Ulrich A. W. Tetzlaff, Erwin Pesch, Optimal Workload Allocation between a Job Shop and an FMS, IEEE TRANSACTIONS ON ROBOTICS AND AUTOMATION, VOL. 15, NO. 1, FEBRUARY 1999, 20-32

5. Mansour Abou Gamila, Saeid Motavalli, A modelling technique for loading and scheduling problems in FMS, Robotics and Computer Integrated Manufacturing 19 (2003) 45–54

6. Mohit Goswami, M. K.Tiwari , S. K. Mukhopadhyay, An integrated approach to solve tool-part grouping, job allocation and scheduling problems in a flexible manufacturing system, Int. J. Adv. Manuf. Technol. DOI 10.1007/s00170-006-0796-8

7. Xiao ming ZHANG, Susumu FUJI, Toshiya KAIHARA, Evaluation of Tool Allocation Strategies in Flexible Manufacturing System, JSME Int. J., Ser. C, Vol. 48, No. 1, (2005),pp.37-45

8. S. Rahimifard, S.T. Newman, Machine loading algorithms for the elimination of tardy jobs in flexible batch machining applications, Journal of Materials Processing Technology 107 (2000) 450-458

9. Chinyao Low, Yukling Yip, Tai-Hsi Wu. Modelling and heuristics of FMS scheduling with multiple objectives. Computers & Operations Research 33 (2006) 674–694

10. L. J. Zeballos, O. D. Quiroga, G. P. Henning, A constraint programming model for the scheduling of flexible manufacturing systems with machine and tool limitations, Engineering Applications of Artificial Intelligence 23 (2010) 229–248

11. F. Jolai, H. Asefi, M. Rabiee, P. Ramezani, Bi-objective simulated annealing approaches for no-wait two-stage flexible flow shop scheduling problem, Scientia Iranica E (2013) 20 (3), 861–872

12. Adam S. Sikor, Higher Mathematics, Springer, 2012, p20, (Theorem 7)
16